\documentclass[12pt,preprint]{elsarticle}

\usepackage{amssymb,amsmath}

 \usepackage{array,multirow}
 \usepackage{graphicx}
 \usepackage{caption}
 \usepackage{hyperref}
 \usepackage{booktabs}
 \usepackage{color,soul}

\newcommand{\bftab}{\fontseries{b}\selectfont}

\graphicspath{ {UMB_figures/} }

\journal{Ultrasound in Medicine and Biology}

\begin{document}

\begin{frontmatter}



\title{Heterogeneous tissue characterization using ultrasound: a comparison of fractal analysis backscatter models on liver tumors}





\author[Affil1,Affil2]{Omar S. Al-Kadi \corref{cor1}}
\author[Affil1]{Daniel YF Chung}
\author[Affil1]{Constantin C Coussios}
\author[Affil1]{J Alison Noble}
\address[Affil1]{Institute of Biomedical Engineering, Department of Engineering Science, University of Oxford, Oxford, UK}
\address[Affil2]{King Abdullah II School for Information Technology, University of Jordan, Amman, Jordan}
\cortext[cor1]{Corresponding Author: E-mail. o.alkadi@ju.edu.jo.}

\begin{abstract}
Assessing tumor tissue heterogeneity via ultrasound has recently been suggested for predicting early response to treatment. The ultrasound backscattering characteristics can assist in better understanding the tumor texture by highlighting local concentration and spatial arrangement of tissue scatterers. However, it is challenging to quantify the various tissue heterogeneities ranging from fine-to-coarse of the echo envelope peaks in tumor texture. Local parametric fractal features extracted via maximum likelihood estimation from five well-known statistical model families are evaluated for the purpose of ultrasound tissue characterization. The fractal dimension (self-similarity measure) was used to characterize the spatial distribution of scatterers, while the Lacunarity (sparsity measure) was applied to determine scatterer number density. Performance was assessed based on 608 cross-sectional clinical ultrasound radio-frequency images of liver tumors (230 and 378 demonstrating respondent and non-respondent cases, respectively). Cross-validation via leave-one-tumor-out and with different \textit{k}-folds methodologies using a Bayesian classifier were employed for validation. The fractal properties of the backscattered echoes based on the Nakagami model (Nkg) and its extend four-parameter Nakagami-generalized inverse Gaussian (NIG) distribution achieved best results -- with nearly similar performance -- for characterizing liver tumor tissue. Accuracy, sensitivity and specificity for the Nkg/NIG were: 85.6\%/86.3\%, 94.0\%/96.0\%, and 73.0\%/71.0\%, respectively. Other statistical models, such as the Rician, Rayleigh, and K-distribution were found to not be as effective in characterizing the subtle changes in tissue texture as an indication of response to treatment. Employing the most relevant and practical statistical model could have potential consequences for the design of an early and effective clinical therapy.   
\end{abstract}


\end{frontmatter}

\pagebreak









\section*{Introduction}
\label{intro}
Ultrasound tissue characterization can provide useful quantitative assessments for understanding the state of biological disease \citep{mao13}. With advancement in medical image analysis, it is becoming a promising non-invasive technique for early detection of tumor response to treatment \citep{cza13,sdg13_ASA,sdg13_CCR}. It has the advantage of deriving parameters that can represent tissue properties in a fast, non-ionizing, easily operated, and cost-effective way compared to other conventional follow-up imaging techniques. Soft tissue pathologies in the form of lesions tend to have distinct scattering patterns to that of normal tissue structure, and the associated acoustic properties could characterize the concentration of scatterers and micro-structures; which is an indication of different tissue types. Biological tissue ultrasonic modeling followed by echo signal analysis can facilitate heterogeneity examination of tumor texture.

The interaction of an acoustic wave with different tissue regions can be modeled by the backscattered radio-frequency (RF) signal. Tissue properties based on the scatterer number density and spatial distribution can be derived subsequently for analysis. There are several approaches for which useful information can be extracted from the RF signal. One approach uses the local power spectral density to estimate the integrated backscatter and attenuation coefficient \citep{sch14,rbt14,nam11,bsh08}, or to measure the mean central frequency and scatterer size \citep{bdl97,nbg15,lvo12,sha11}. Textural properties of the tissue spatial arrangement can also be estimated from the envelope-detected RF image \citep{bhl09,kln11}. As the first-order statistical properties of the backscattered RF signal rely on the number density and spatial distribution of scatterers \citep{wgr83,pra12} -- which maybe coherent, random or a mixture of both -- it would be difficult to account for all scatterer conditions using the former approaches. Therefore others have investigated the probability density function of the backscattered echoes and proposed to account for the number, size, spacing and regularity of the scatterers in tissue. An overview of the various statistical distributions for modeling the envelope-detected RF signal can be found in \citep{dsp10}. For the latter approach, the main objective is to provide a better characterization of the fundamental elements that form the coarse textural patterns, namely speckles formed from the backscattered echoes. The speckle local arrangement represents the various scatterer concentrations and spatial distributions occurring in tissue; ranging from a fully developed, partially developed, to a coherent speckle pattern. In cases when there are many randomly located scatterers per resolution cell (i.e. fully developed speckle), the envelope signal statistics would follow a square root of exponential distribution, known as Rayleigh distribution \citep{wgr83}. The model can be further subdivided into pre-, Rayleigh, and post-Rayleigh for characterizing heterogeneous, homogeneous, and periodic textures, respectively \citep{cbt99,skr96,mln95}. Non-Rayleigh distributions can be observed when the scatterers become less condensed or having a structure with some regularity. That is, when the number of scatterers in a resolution cell is small (i.e. partially developed speckle), the K-distribution was found to be more effective in modeling the pre-Rayleigh statistics of the RF envelope \citep{skr00_UMB}. The RF envelope statistics in this case would have a shape resembling a square root of the product of a gamma and exponential distributions. Whereas if the scatterers become organized with some periodicity (i.e. coherent speckle), the Rician distribution -- which is the generalization of the Rayleigh distribution -- is more appropriate for characterizing the underlying regular structures in tissue \citep{sjb98}.

Although the prior models account for specific aspects of scatterer localization, they are not general enough to model the various tissue texture conditions. This is true when a high degree of variability in the scattering cross-section associated with a low number of scatterers exist. In practice it is very common to encounter non-Rayleigh conditions of ultrasound backscatter, such as mixtures of diffuse and coherent or periodically aligned scatterers in tissue micro-structure. Therefore a comprehensive approach is sought as in the generalized K-distribution \citep{Jkm87} and homodyne K-distribution \citep{dut94,mao11}. A third parameter, which represents the envelope of the coherent signal, was added to the effective number of scatterers and energy of the random scatterers to account for post-Rayleigh conditions. However a drawback lies in the analytical complexity of these model generalizations, rendering the process of parameter estimation computationally expensive. Other models which also apply a generalization approach to the Rayleigh and Rice distributions to better-fit the backscattered echo include: Weibull \citep{rju02}, Rician inverse Gaussian \citep{etf05}, and generalized gamma \citep{tns05}. On the other hand, the Nakagami distribution family can provide a simpler and general model for ultrasonic tissue characterization \citep{skr00_UFFC}. The regularity of the scatterer spacing are taken into consideration besides the scatterer density and amplitude, making it possible to account for hypoechoic and hyperechoic structures \citep{tsi10}. The model shape equivalent to a scaled square root of a gamma distribution can better represent the backscattered RF signal envelope, and can be easily fine-tuned via the shape Nakagami parameter to represent low and high number of scatterer densities with minimal error \citep{tsi14}. The model was further generalized as a Nakagami-generalized inverse Gaussian distribution in \citep{krm06} by including an additional shape adjustment parameter to account for the tails of the density function. However this generalization was not investigated with real tissue, where scatterers tend to have a high degree of variability in scattering cross-sections. 

All aforementioned statistical models of the backscattered echo envelope in the literature claim a better characterization of texture anisotropic properties. A large number of articles address the problem of soft tissue characterization and diagnosis from ultrasound images of various internal organs, such as in kidneys \citep{wu13}, liver \citep{ghl12}, breast \citep{tdn14}, gallbladder \citep{kmn10}, pancreas \citep{ate14}, spleen \citep{rsn13} and abdominal aorta \citep{tsi08} all of which demonstrates practical examples, but not limited to, of recent clinical work . However, to the best of our knowledge, the study of the performance of the most well-known statistical models for characterizing ultrasonic liver tumor tissue has not been investigated. Local parametric fractal features extracted via maximum likelihood estimation for each statistical model will be used for liver tumor texture classifications. In this paper a fractal approach for evaluating the performance of the different statistical models would be sought for an automated detection of liver tumor response to chemotherapy treatment. The approach relates the fractal characteristics of the tissue scatterers to the underlying statistical properties derived from the RF envelope-detected signal. The fractal dimension which represents the degree of self-similarity and the derived Lacunarity which indicates the level of spaces within the texture were related to the scatterers spatial distribution and number density, respectively. The fractal features are extracted via a multimodal statistical distribution and results fed to a classifier for automated classification. The aim of the paper is to demonstrate the efficacy in modeling the underlying tumor physiological changes from ultrasound parametric images of scatterer tissue properties. Tumor tissue scatterer characterization is a challenging task due to the disease chaotic behavior. Therefore we will address the issue of best model selection, and whether analytically complex models are really necessary for better characterizing complex tissue, such as liver tumors. 


The paper is organized as follows: the preference of working with RF envelope-detected signal rather than B-mode images is briefly explained in the ultrasound data representation section, and followed by modeling the backscattered RF signal to cover the different scatterer conditions and densities. Then the multiscale feature extraction section discusses characterizing the fractal patterns of the tissue scatterers for automated classification. The results and discussion sections list and discuss the major findings, and the paper is summarized in the conclusion section.

\section*{Ultrasound data representation}
\label{MaM}
      
\subsection*{RF parametric image acquisition}
The acquired ultrasound beams run through several signal and image pre-processing steps before they can be presented as a radio-frequency (RF) matrix, and subsequently in a gray scale B-mode image, see Fig.1.
The ultrasound operator may also choose to further optimize the visibility by increasing the gain of the reflected signals with increasing time from the transmitted pulse (i.e. time-gain compensation).
While the B-mode scans could be suitable for qualitative analysis in describing the tissue structure, some aspects could be in fact obscured or modified, and would not realistically represent the condition in the raw data. The post-processing functions that have been applied manipulate the image on the screen, without improving the signal quality itself, or the fundamental signal to noise ratio. 
The RF envelope-detected signal can better provide quantitative analysis and balance the tradeoff between preserving most of the ultrasound data unaltered and with minimal filtering (i.e. only removal of high frequency oscillations). Many works in the literature has shown the usefulness of RF data for clinical analysis as mentioned in Section 1, therefore this work utilizes the envelope-detected RF data for ultrasound tissue characterization of liver tumors. 

\begin{figure} [ht]
\centering
\includegraphics[scale = 7.0]{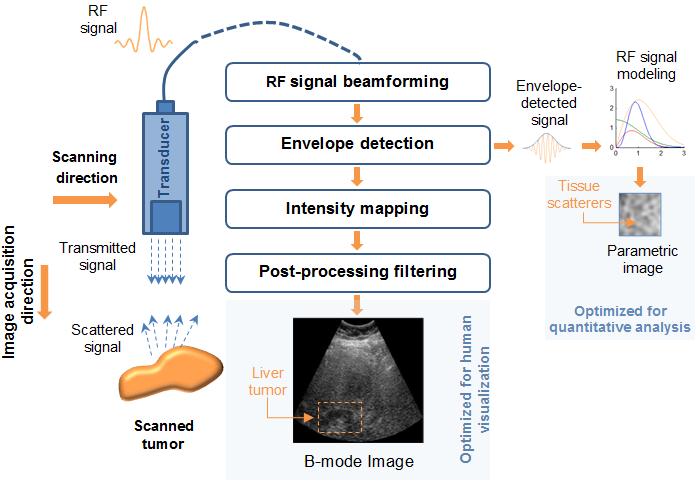}
\caption{Illustration of ultrasound RF to B-mode data conversion process.}
\label{fig:RF_to_B-mode}
\end{figure}

\subsection*{Clinical liver tumor cross-sectional images}      

Cross-sectional images of liver tumors undergoing chemotherapy treatment obtained as part of an ethically approved prospective study was used to validate the clinical significance of the different statistical models. A total of 608 cross-sectional images (230 from tumors responded to treatment categorized as respondent acquired from 11 patients, and 378 from tumors progressed categorized as non-respondent acquired from 22 patients) were obtained using a diagnostic ultrasound system (z.one, Zonare Medical Systems, Mountain View, CA, USA) with a 4 MHz curvilinear transducer and 11 MHz sampling. 

The target tumor was initially imaged using the optimal acoustic window as determined by an experienced operator with 5 years experience of liver ultrasound imaging. The depth of imaging was kept constant for all study acquisitions. Other imaging settings including scanning frequency, gain and time gain compensation were also kept constant. The operator then positioned the imaging transducer to ensure the target tumor is optimally placed at the center of the image. While maintaining stable skin contact position, the transducer was angled to image the most cranial aspect of the tumor. The transducer was then swept in a fan motion towards the caudal aspect of the tumor, capturing multiple cross sectional images through the target tumor. The procedure was repeated at all subsequent time points during the monitoring period. Since the imaging protocol was constant, the tumor should be at identical depth every time it is imaged throughout the monitoring period. 

Each dataset was acquired prior to commencement of chemotherapy. Response to treatment was determined based on conventional computed tomography follow up imaging as part of the patient standard clinical care based on the response evaluation criteria in solid tumors (RECIST) \citep{eis09}. The baseline cross-sectional imaging was compared against those performed at the end of treatment according to the RECIST criteria to determine response to treatment for each target tumor. A tumor was classified as responsive if categorized as partial response and non-responsive if no change or disease demonstrated progression. Cross-validation was performed based on a leave-one-tumor-out approach for a total of 40 volumetric (stacks of segmented 2D slices) tumors, for which 13 were responsive and 27 did not respond to chemotherapy treatment. Imaging was done by two radiologists to provide gold standard and training data. The acquisition of the dataset in this work did not influence the diagnostic process or the patients' treatment, and informed consent was received from each participant in the study.

\section*{Modeling the backscattered RF signal}

Ultrasound backscattering in tissue can be considered as a random process. Many statistical models have been applied for describing the probability density function of the envelope-detected RF signal. Evaluation has been made by five major statistical distributions that are widely used in the literature to describe the various aspects of tissue characteristics.

\subsection*{Rayleigh distribution}
The Rayleigh distribution ideally suits the condition when the amplitudes of the individual backscattered signals are assumed to be randomly distributed, i.e. speckle pattern is ``fully developed''. The \textit{n}-dimensional Rayleigh distribution \citep{dsp10} can be seen as a continuous probability distribution for positive-valued random variables defined 
\begin{equation}
P_{ray}\left(x|\sigma^2\right) = \frac{2}{\Gamma\left(n/2\right)}\left(\frac{1}{2\sigma^2}\right)^{n/2} x^{n-1} e^{-x^2/2\sigma^2},
\end{equation}
where $x$ represents the amplitude of the signal, $\sigma > 0$ is the scale parameter of the distribution, and $\Gamma$ denotes the Euler gamma function. The case when $n = 2$ corresponds to the Rayleigh distribution.

\subsection*{Rician distribution}
The Rician or Rice distribution model combines a coherent component to the backscattered signal that underlies the Rayleigh distribution. It has the capability of modeling unresolved structures exhibiting a periodic pattern in the spatial scatterer distribution \citep{ukr09}. An \textit{n}-dimensional Rician distribution \citep{sjb98} can be expressed as
\begin{equation}
P_{ric}\left(x|\epsilon,\sigma^2\right) = \left(\frac{\epsilon}{\sigma^2}\right) \times \left(\frac{x}{\epsilon}\right)^{n/2} I_{n/2-1}\left(\frac{\epsilon}{\sigma^2}x\right) e^{-\left(\epsilon^2 + x^2\right)/2\sigma^2},
\end{equation}
where $x$ represents the amplitude of the signal, the scale parameter $\sigma > 0$ and non-centrality parameter $\epsilon \geq 0$ are real numbers, $n$ is the dimension and $I_{p}$ denote the modified Bessel function of the first kind of order $p$. When $n = 2$ the distribution would correspond to a Rician distribution.
The special case where $\epsilon \rightarrow 0$ would yield the Rayleigh distribution.

\subsection*{K-distribution}
The K-distribution can be considered as a generalization of the Rayleigh distribution that can account for non-uniformities in the spatial scatterer distribution. This property allows the modeling of clusters of scatterers or when the effective number of scatterers per resolution cell is small. The K-distribution \citep{dsp10} is defined by
\begin{equation}
P_{kd}\left(x|\sigma^2,\alpha\right) = \frac{4x^{\alpha-1+n/2}}{\left(2\sigma^2\right)^{\left(\alpha+n/2\right)/2}\Gamma\left(\alpha\right)\Gamma\left(n/2\right)}K_{\alpha-n/2}\left(\sqrt{\frac{2}{\sigma^2}}x\right),
\end{equation}
where $\alpha > 0$, $\sigma^2 > 0$ are the shape and scale parameters, respectively, and $K_p$ denotes the modified Bessel function of the second kind of order $p$. The K-distribution includes a Rayleigh distribution as a special case for $\alpha = \infty$.

\subsection*{Nakagami distribution}
This is another family of distributions that can model the ultrasonic backscattered envelope for various scattering conditions and scatterer densities. The Nakagami density function \citep{nka60} is defined as
\begin{equation}
P_{nkg}\left(x|m,\Gamma\right) = \frac{2m^m}{\Gamma\left(m\right)\Omega^m}x^{2m-1}e^{-mx^{2}/\Omega},
\end{equation}
for $x \geq 0$, where $\Gamma$ is the Euler gamma function. The real numbers $m > 0$ (related to the local backscattered energy) and $\Omega > 0$ (related to the local concentration of scatterers) are called the shape and scaling parameters, respectively. Similarly to the Rayleigh distribution, the envelope of the RF signal $x^2$ follows a gamma distribution. By fine-tuning the shape of the distribution parameter $m$, other statistical distributions can be modeled, such as, an approximation of the  Rician distribution (i.e. post-Rayleigh) for $m > 1$, a Rayleigh distribution for the special case when $m = 1$, and when $m < 1$ a K-distribution (i.e. post-Rayleigh) for modeling the backscattered echoes with less dense scatterers. Thus the SNR of the Nakagami distribution can take any positive value.

\subsection*{Nakagami-generalized inverse Gaussian distribution}
Based on a four parameter model, the Nakagami-generalized inverse of Gaussian distribution (NIG) distribution \citep{krm06,dsp10} is an extended form of the Nakagami distribution to better cover the coherent scattering conditions. It is defined by
\begin{align} 
P_{NIG}\left(x|m,\Omega,\theta,\lambda\right) = \frac{2\left(m/\Omega\right)^m}{\lambda^{\theta/2}K_{\theta}\left(\sqrt{\lambda}\right)\Gamma\left(m\right)}x^{2m-1} \nonumber \\ \times \left(\frac{\Omega}{2mx^{2}+\lambda\Omega}\right)^{\left(m-\theta\right)/2} K_{\theta-m}\left(\sqrt{\lambda+\frac{2mx^2}{\Omega}}\right),
\end{align}
where $m$, $\Omega$, $\theta$ and $\lambda$ are positive real numbers with the condition $\theta > 0, \lambda \geq 0$, and $K$ is modified Bessel function of the second kind. By having the four parameter values of the NIG distribution model set to specific conditions, various well-known distributions can be recovered such as, Rayleigh ($m = 1, \theta > 0, \lambda\Omega \rightarrow \infty$), Nakagami ($m > 1, \theta > 0, \lambda\Omega \rightarrow \infty$), K-distribution ($m = 1, \theta > 0, \lambda\Omega \rightarrow 0$), and Nakagami-gamma ($m > 0, \theta > 0, \lambda\Omega \rightarrow 0$).

Besides the NIG distribution, other complex models such as the Generalized-K, the Homodyned-K, and the Rician inverse of Gaussian distributions exist in the literature for ultrasound tissue characterization. However they were not used as the simpler and more general models of Nakagami family -- in its 2 and 4-parameter extended form -- which under different limiting conditions can approximate the known distributions, and hence can cope with the various scattering conditions related to the coherent-to-diffuse signal power ratio and scatterer clustering, see illustration in Fig.2. 

\begin{figure} [ht]
\centering
\textbf{Scatterer characterization}
\includegraphics[scale = 1]{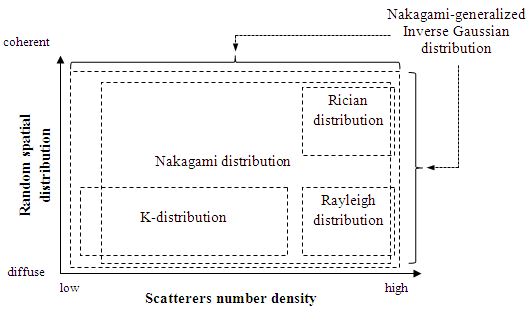}
\caption{An illustration representing the coverage of the scatterer distribution vs scatterer number density by the 5 statistical distributions employed in this work for tissue characterization.}
\label{fig:scatterer_characterization}
\end{figure}
		
\section*{Multiscale feature extraction}

\subsection*{Parametric image generation}
Based on a voxel-by-voxel approach, the statistical model parameters are estimated from the RF envelope data. Each estimated parameter in the generated parametric image corresponds to a voxel -- short segment -- of the RF signal. As we are dealing with a tumor volume, the RF envelope signal was presented as a matrix, and several matrices represent the tumor volume. Then the different statistical distribution models are fitted and then their associated model parameters estimated by maximum likelihood estimation. Namely, the maximum likelihood estimate $\hat{\theta} \left(v\right)$ for a density function $f\left(v_{1},\ldots,v_{m}/\theta\right)$ when $\theta$ is a vector of parameters for a certain distribution family $\Theta$, estimates the most probable parameters $\hat{\theta}\left(v\right) = arg max_\theta \: D\left(\theta/v_{1},\ldots,v_{m}\right)$, where $D\left(\theta /v\right) = f\left(v/\theta\right), \theta \in \Theta$ is the score function. Thereby, a number of parametric images $l_r$, where $r = 1, 2,$ or $4$ for a mono, bi, and quad-parametric model, respectively, are generated for each corresponding RF envelope image of the applied distribution model. Herein, $l_r$ refers to the statistical model parameters estimated locally from the neighborhood of the RF envelope data by a square--sliding window with its size optimized as in \citep{alk14b} and centered on each point in the RF data matrix. Also the same RF envelope data was used to create the different parametric images for each statistical model. That is, creating for each model maps of the distribution parameters throughout the entire scanned region (i.e. each parameter map being an image). 

An interesting point to consider is the difference between the spatial distribution of tissue scatterers, and the structure of the texture of the parametric ultrasound images. The former results from the interaction of the sound energy with the liver tumor tissue leading to scatter in many directions. Complex structure of tissue scatterers and their small size as compared to the ultrasound wavelength, gives the speckle patterns a Rayleigh scattering behavior. The spatial distribution of these tissue scatterers is not related to any particular ultrasound imaging system, but is a property of the tissue structure and the ultrasound frequency used which dictates the scattering length scale. On the other hand, texture of the parametric ultrasound images, which shows different spatial variations of speckle intensity, is dependent on the used ultrasound imaging system. Usually a set of proprietary filters -- specific to the ultrasound device manufacturer -- are used for post-processing operations, which may result in loss of information. Therefore the RF envelope-detected data in this work did not undergo any post-processing filtering or time-gain compensation operations.

\subsection*{Fractal features}
Various statistical signatures can be derived from the acoustic properties of the backscattered RF and envelope-detected signals. An effective way to investigate the local arrangement of the scatterer concentrations and spatial distributions occurring in tissue is using a multifractal analysis approach. The fractal dimension (FD) can indicate the degree of ``self-similarity'' in the spatial distribution of tissue scatterers, and has shown many promising results in tissue characterization from different imaging modalities \citep{alk08, alk14a, alk14b, mvk06, ifd03}. Also the Lacunarity, which is a measure derived from the texture of the FD, reflects the “gaps” within the tissue texture, and hence measures the number density of tissue scatterers. Both signatures, the FD defined in (\ref{eq:FD}) -- estimated via the fractal Brownian motion which is a non-stationary model known for its capability for describing random phenomena \cite{lop09} -- and its counterpart Lacunarity which is the standard deviation divided by the mean of a fractal image $F_p$ using a sliding box-counting algorithm as shown in (\ref{eq:Lac}), are considered useful for revealing the mixtures of spatial homogeneity and heterogeneity properties of tissue texture in the generated parametric images. 

\begin{equation}
FD = \frac{log\left(N_s\right)}{log\left(1/s\right)},
\label{eq:FD}
\end{equation}
\begin{equation}
L = \frac{1/MN\sum^{M-1}_{m=0}\sum^{N-1}_{n=0}F_p\left(m,n\right)^2}{\left(1/MN\sum^{M-1}_{k=0}\sum^{N-1}_{l=0}F_p\left(k,l\right)\right)^2} - 1
\label{eq:Lac}
\end{equation}
where $N_s$ is the number of self-similar shapes and $s$ is the corresponding scaling factor; $M$ and $N$ in (\ref{eq:Lac}) are the size of the fractal parametric image $F_p$.

In this work, the FD and derived Lacunarity signatures are estimated in a multiscale fashion after applying a Daubechies wavelet packet analysis decomposition technique, where the texture variations of the subbands at each level of decomposition serve as the extracted fractal feature pattern. The Daubechies wavelet can account for self-similarity and signal discontinuities, making it useful for characterizing signals exhibiting fractal patterns \citep{dub90}. An orthogonal 8-tap Daubechies filter was used to obtain the wavelet packets by expanding the basis having the most significant fractal signature rather than energy \citep{alk09_2}. The fractal features at the different resolution scales give more information on the local concentration and spatial arrangement of the tissue scatterers -- as higher FD signature values indicate a more heterogeneous texture (i.e. distribution of scatterers are random), and vice versa \citep{dav12,mtr09}; whereas Lacunarity signature values operate oppositely, giving a low value when the scatterers are large, viz. less gaps between scatterers per resolution cell, see Fig.3. An illustrative example of liver tumor tissue characterizing by the different statistical models is shown in Fig.4. The first column shows the distribution and number density of the tissue scatterers modeled by the different statistical distributions, while the second column represents the textural patterns of the modeled tissue scatterers in a multifractal analysis approach.

At the end of the feature extraction stage, a feature vector $\Im = (w^{l}_{1,1}, w^{l}_{1,2},$\\$ \ldots, w^{l}_{r,q})$ would consist of all selected fractal signatures $w$ for a specific model parameter $r$ and wavelet subband $q$ from each RF envelope image $l$. A data pipeline analysis of the fractal feature extraction process is illustrated in Fig.5.

\begin{figure}
\centering
\textbf{Fractal characterization}
\includegraphics[scale = 11]{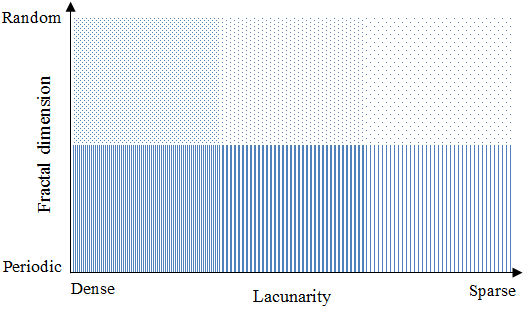}
\caption{Tissue characterization using the fractal properties of the scatterer’s distribution and clustering: The fractal dimension measures the level of self-similarity (i.e. spatial distribution), while the Lacunarity assesses the degree of heterogeneity (i.e. denseness) in tissue texture.}
\label{fig:fractal_characterization}
\end{figure}

\begin{figure}
\centering
\includegraphics[scale = 7]{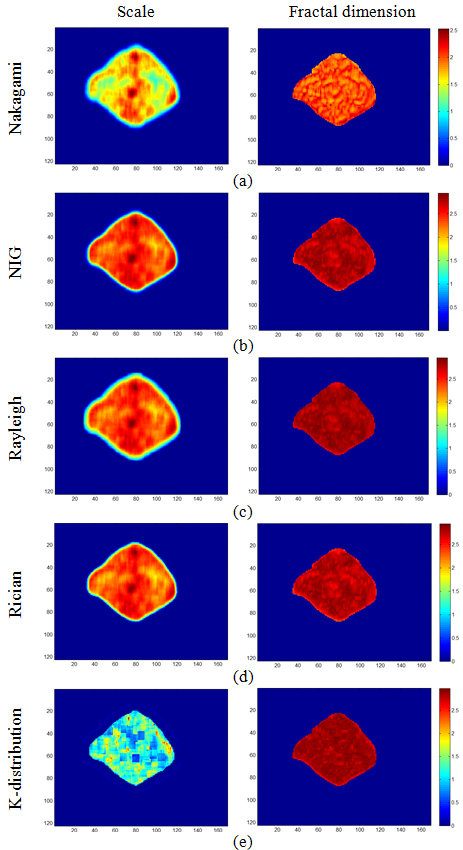}
\caption{Characterizing a segment of a liver tumor tissue: (from left to right) model parametric image and corresponding fractal image for (a) Nakagami, (b) Nakagami-generalized inverse Gaussian, (c) Rayleigh, (d) Rician, and (e) K-distribution statistical models, respectively. }
\label{fig:model_vs_fractal_characterization}
\end{figure}

\begin{figure}
\centering
\includegraphics[scale = 0.75]{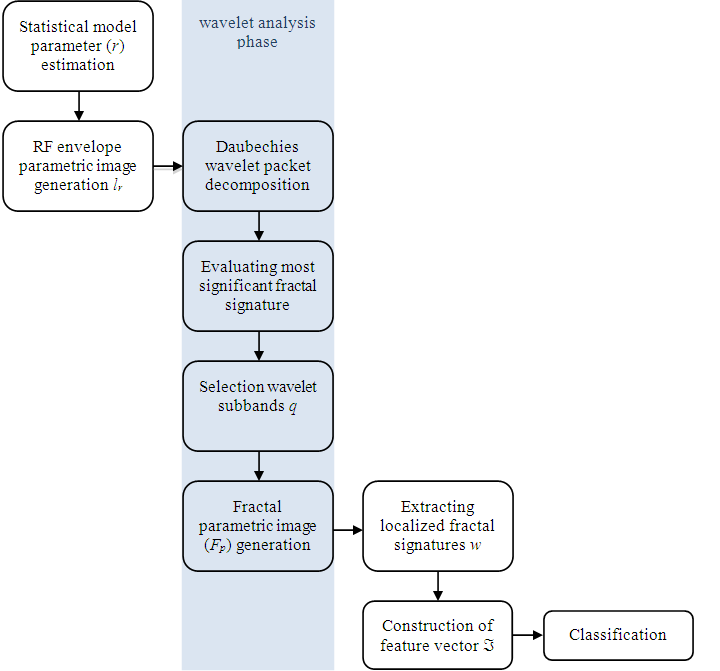}
\caption{Data pipeline analysis explaining the fractal feature extraction process.}
\label{fig:pipeline_analysis}
\end{figure}
\subsection*{Pattern classification}
In order to assess the quality of the extracted features from the different statistical distribution models, a na\"{\i}ve Bayesian classifier (NBC) was used for classification. The NBC is a robust learning method which can work well even when the data is non-normally distributed \citep{IR08}, and the assumption of attribute independence allows the NBC classifier to estimate the classification accuracy using less training data as compared to other classifiers. This fast and simple classifier was reported to perform well in practice even with the presence of strong attribute dependence \citep{dmg97}. According to Bayesian theory, each fractal feature $\Im_f, f = 1, \ldots, n$, would be assigned to class $i$, if $P\left(C_i|\Im\right)$ is maximized over all classes $P\left(C_i|\Im\right) > P\left(C_j|\Im\right) \forall j \neq i$.

To find $P\left(C_i|\Im\right)$, NBC can be represented as:

\begin{equation}
P\left(C_i|\Im\right) = \left(\frac{P\left(C_i\right)\prod^{n}_{\substack{f=1}}P\left(\Im_f|C_i\right)}{\sum^{k}_{i=1}P\left(C_i\right)\prod^{n}_{\substack{f=1}}P\left(\Im_f|C_i\right)}\right)
\label{eq:NBC}
\end{equation}
where $P\left(C_i|\Im\right)$ is the a $\textit{posteriori}$ probability of assigning class $i$ given feature vector $\Im$, $P\left(\Im_f|C_i\right)$ is the probability density function of $\Im$ within the $i^{th}$ class $C_i$ for a total number of $k$ classes, $P\left(C_i\right)$ and $P\left(\Im\right)$ are the a $\textit{priori}$ probability of class $C_i$ and feature vector $\Im$; respectively.       

\section*{Results}
\label{Results}
The extracted fractal features from the texture of the various parametric images -- which where derived from the RF envelope-detected signal of the different distribution models --are compared in Tables 1, 2, and 3, where values in bold indicate the best achieved performance. A detailed classification performance of the 3D clinical ultrasound liver tumor test set via a leave-one-out cross-validation approach is shown in Table 1. The tissue characterization based on the Nakagami (Nkg) distribution and its 4-parameter extended version (NIG distribution) outperformed the other statistical models in terms of classification performance. Both Nkg and NIG had nearly equal classification accuracy and the area under the receiver operating characteristics (ROC) curve with 85.6\% \& 86.3\%, and 83.2\% \& 83.5\%, respectively. Although NIG had the highest sensitivity (true-positive rate), the Nkg achieved relatively better in the statistical measures of specificity (true-negative rate), precision, and false-positive (FP) rate. The classification performance of the Rayleigh distribution was the third among all five distributions; nevertheless achieving equal statistical measures regarding specificity and FP rate with the NIG model. The Dice similarity coefficient affirms the accuracy results and shows how similar the predicted and observed responses are in regard to chemotherapy treatment. Although the K-distribution was the least sensitive in terms of classification performance -- next to Rician distribution, it recorded the best specificity score as compared to the other distributions.

In order to avoid overfitting and to give some insights on how would the different distribution models based on the scatterers fractal characteristics generalize, the extracted features were further validated. Table 2 and 3 show the detailed classifier performance on the test set for 10- and 20-fold cross-validation (results are the mean $\pm$ standard deviation of the performance over 60 runs), respectively.

\begin{table} [!ht]
\centering
\captionsetup{justification=centering}
\caption{Detailed Classification Performance (leave-one-out cross-validation) for different statistical models of RF ultrasound for liver tissue characterization}
\label{table:fractal-based_loo}
\begin{tabular}{lccccc}
\hline \hline
\multirow{2}{*}{\begin{tabular}[c]{@{}c@{}}Classification \\ Performance\end{tabular}} & \multicolumn{5}{c}{RF-based stochastic distribution}   \\ \cline{2-6}
 & Nkg & Ric & Ray & NIG & Kd   \\ \hline
FP rate & \bftab 0.271 & 0.327 & 0.289 & 0.289 & 0.336 \\
Sensitivity  & 0.940 & 0.890 & 0.920 & \bftab 0.960 & 0.880 \\
Specificity  & 0.730 & 0.670 & 0.710 & 0.710 & \bftab 0.770 \\
Accuracy & 0.856 & 0.810 & 0.838 & \bftab 0.863 & 0.798 \\
Precision & \bftab 0.850 & 0.810 & 0.830 & 0.840 & 0.795 \\
Dice SC & 0.922 & 0.892 & 0.912 & \bftab 0.926 & 0.888 \\
ROC Area & 0.832 & 0.781 & 0.814 & \bftab 0.835 & 0.773 \\ \hline \hline
\end{tabular}
\end{table}

\begin{table} [!ht]
\centering
\captionsetup{justification=centering}
\caption{Detailed Classification Performance (10-fold cross-validation) for different statistical models of RF ultrasound for liver tissue characterization}
\label{table:fractal-based_10-fold}
\hspace*{-1.7cm}
\begin{tabular}{lccccc}
\hline \hline
\multirow{2}{*}{\begin{tabular}[c]{@{}c@{}}Classification \\ Performance\end{tabular}} & \multicolumn{5}{c}{RF-based stochastic distribution}   \\ \cline{2-6}
 & Nkg & Ric & Ray & NIG & Kd   \\ \hline
    FP rate & \bftab 0.280 $\pm$ \bftab 0.076 & 0.345 $\pm$ 0.113 & 0.285 $\pm$ 0.089 & 0.332 $\pm$ 0.045 & 0.380 $\pm$ 0.128  \\
Sensitivity & 0.924 $\pm$ 0.720 & 0.887 $\pm$ 0.655 & 0.911 $\pm$ 0.715 & \bftab 0.955 $\pm$ \bftab 0.668 & 0.870 $\pm$ 0.619  \\
Specificity & \bftab 0.720 $\pm$ \bftab 0.924 & 0.655 $\pm$ 0.887 & 0.715 $\pm$ 0.911 & 0.668 $\pm$ 0.955 & 0.620 $\pm$ 0.872  \\
Accuracy & \bftab 0.845 $\pm$ \bftab 0.007 & 0.797 $\pm$ 0.009 & 0.836 $\pm$ 0.007 & \bftab 0.845 $\pm$ \bftab 0.009 & 0.780 $\pm$ 0.011  \\
Precision & \bftab 0.840 $\pm$ \bftab 0.856 & 0.803 $\pm$ 0.785 & 0.836 $\pm$ 0.836 & 0.821 $\pm$ 0.905 & 0.790 $\pm$ 0.754  \\
Dice SC & \bftab 0.916 $\pm$ \bftab 0.004 & 0.887 $\pm$ 0.005 & 0.910 $\pm$ 0.004 & \bftab 0.916 $\pm$ \bftab 0.006 & 0.870 $\pm$ 0.007  \\
ROC Area & \bftab 0.822 $\pm$ \bftab 0.009 & 0.771 $\pm$ 0.010 & 0.813 $\pm$ 0.008 & 0.812 $\pm$ 0.011 & 0.75 $\pm$ 0.012  \\ \hline \hline
\end{tabular}
\hspace*{-1.7cm}
\end{table}

\begin{table} [!ht]
\centering
\captionsetup{justification=centering}
\caption{Detailed Classification Performance (20-fold cross-validation) for different statistical models of RF ultrasound for liver tissue characterization}
\label{table:fractal-based_20-fold}
\hspace*{-1.7cm}
\begin{tabular}{lccccc}
\hline \hline
\multirow{2}{*}{\begin{tabular}[c]{@{}c@{}}Classification \\ Performance\end{tabular}} & \multicolumn{5}{c}{RF-based stochastic distribution}   \\ \cline{2-6}
 & Nkg & Ric & Ray & NIG & Kd   \\ \hline
FP rate & \bftab 0.274 $\pm$ \bftab 0.073 & 0.332 $\pm$ 0.112 & 0.283 $\pm$ 0.084 & 0.308 $\pm$ 0.041 & 0.361 $\pm$ 0.124  \\
Sensitivity & 0.927 $\pm$ 0.726 & 0.888 $\pm$ 0.668 & 0.916 $\pm$ 0.717 & \bftab 0.959 $\pm$ \bftab 0.692 & 0.876 $\pm$ 0.639  \\
Specificity & \bftab 0.726 $\pm$ \bftab 0.927 & 0.668 $\pm$ 0.888 & 0.717 $\pm$ 0.916 & 0.692 $\pm$ 0.959 & 0.639 $\pm$ 0.876  \\
Accuracy & 0.849 $\pm$ 0.007 & 0.803 $\pm$ 0.005 & 0.839 $\pm$ 0.007 & \bftab 0.856 $\pm$ \bftab 0.008 & 0.785 $\pm$ 0.008  \\
Precision & \bftab 0.843 $\pm$ \bftab 0.862 & 0.809 $\pm$ 0.789 & 0.837 $\pm$ 0.843 & 0.832 $\pm$ 0.915 & 0.794 $\pm$ 0.765  \\
Dice SC & 0.918 $\pm$ 0.004 & 0.891 $\pm$ 0.003 & 0.912 $\pm$ 0.004 & \bftab 0.922 $\pm$ \bftab 0.004 & 0.879 $\pm$ 0.005  \\
ROC Area & \bftab 0.826 $\pm$ \bftab 0.007 & 0.778 $\pm$ 0.006 & 0.816 $\pm$ 0.008 & \bftab 0.826 $\pm$ \bftab 0.009 & 0.758 $\pm$ 0.010  \\ \hline \hline
\end{tabular}
\hspace*{-1.7cm}
\end{table}

Since we are concerned more with the ability of an early chemotherapy treatment to identify initial tissue response to treatment, namely the sensitivity of the treatment procedure, Fig.6 shows the difference in the sensitivity measure between the raw RF parameters and the fractal-based measures extracted from the parametric images of the RF envelope detected signal. The Wilcoxon signed-rank test on paired overall sensitivity for the fractal and parametric-based of the statistical models shows there is significant differences between the two approaches ($p < 0.05$). For completeness, the models accuracy are shown in Fig.7 as well.

\begin{figure} [!ht]
\centering
\textbf{Sensitivity Comparison}
\includegraphics[scale = 7]{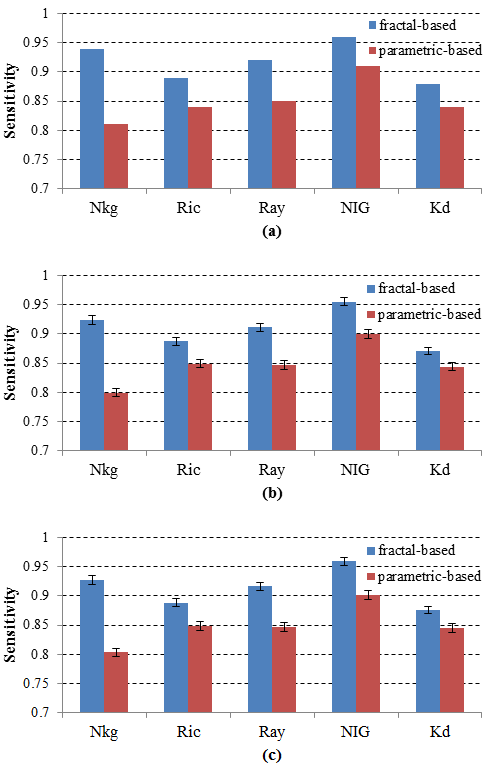}
\caption{Sensitivity comparison of the different statistical models for tissue characterization based on the using the estimated model parameters (parametric-based) and by the fractal properties of the tissue scatterers(fractal-based). The models sensitivity was cross-validated via (a) leave-one-out, (b) 10-fold, and (c) 20-fold.}
\label{fig:comparison_sensitivity}
\end{figure}

\begin{figure} [!ht]
\centering
\textbf{Accuracy Comparison}
\includegraphics[scale = 7]{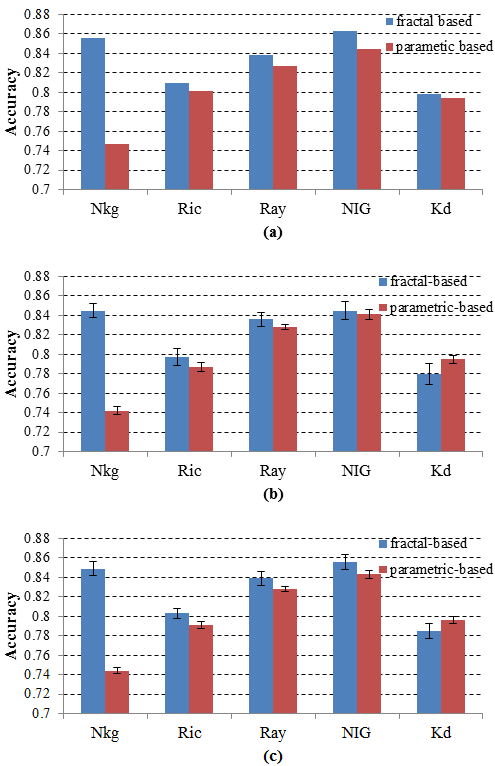}
\caption{Cross-validation accuracy comparison of the different statistical models for tissue characterization based on the using the estimated model parameters (parametric-based) and by the fractal properties of the tissue scatterers(fractal-based). The models accuracy was cross-validated via (a) leave-one-out, (b) 10-fold, and (c) 20-fold.}
\label{fig:comparison_accuracy}
\end{figure}

\section*{Discussion}
\label{Discuss}
Heterogeneity in the tumor tissue scatterers could span different scattering conditions. Regions within the tumor tissue which respond to treatment might exhibit different statistical properties to that of the non-respondent counterpart. Thus it is essential to evaluate the discriminative abilities of the statistical models while simultaneously taking into account all possible scatterer conditions and densities. 

The sum of the individual backscattered signals from the various scattering regions within the biological tissue can be used for RF signal modeling. 
In the case of a fully developed speckle pattern, the size of the scatterers tend to be small in comparison with the wavelength and also have a disorganized structure, i.e. the spatial distribution of scatterers is random \citep{mo92}. This can be considered as a typical case for tumor tissue, where the tumor angiogenesis process causes for the newly induced blood vessels to grow in a chaotic manner, and therefore tissue scatterers to have a random nature. The relation between angiogenesis and randomness of tissue scatterers can be attributed to the chaotic way that tumors build their network of new blood vessels, and thus giving tumorous tissue a ``rougher'' appearance than non-tumorous. These angiogenesis networks tend to be leaky and disorganized, unlike blood vessels in normal tissue. The chaotic behavior introduces a degree of randomness in appearance or ``roughness'' and gives a higher fractal dimension value compared to necrotic tissue. However spatial scatterer distribution in soft tissue is not strictly random and may not originate from identical scatterers, also it is common to encounter different degree of variability in tumor tissue as scatterers vary in shape and size, with a mixture of partially developed and coherent speckle patterns. This compound situation becomes more distinct once certain regions within the tumor start to respond to treatment, which reflects in the statistics of the RF backscattered signal. Moreover, after generating the parametric images from the RF envelope echo, the fractal characteristics of tissue scatterers can provide additional information on the local spatial distribution and number density. Fig.6 and Fig.7 illustrate an improved performance when a fractal characterization is applied to the parametric images. The fractal dimension is analogous to the coherent-diffuse ratio in Fig.2, where it tends to improve distribution separability from a texture perspective by assessing the scatterers random-periodic properties. Also the Lacunarity of the textures in the parametric images further refines the scatterer densities and measures the degree of heterogeneity in tumor tissue. The fractal approach shows a better sensitivity for all distribution models in Fig.6 with the best improvement achieved by the Nakagami model, while the NIG model recorded the highest sensitivity. Similarly in Fig.7, where the accuracy is remarkably improved for the fractal-based images of the Nakagami model, and achieving a nearly comparable performance to that of the complex NIG model. A point to note is the decrease in the K-distribution accuracy for the cross-validations in Fig.7(b) and (c). This indicates that the fractal approach is incapable of extracting useful information from the tissue scatterers characterized by the model, and hence giving inaccurate characterization of tumor heterogeneity. 

In some typical conditions of ultrasound tissue characterization, the backscattered signal cannot be assumed to be Rayleigh distributed, such as when the number of scatterers per resolution cell are not large enough or when there is a lack of randomness in the scatterers localization (e.g. due to periodicity, structure, or clustering) \citep{ukr09}. The former conditions suggest that the Rayleigh distribution may not approximate effectively the statistical characteristics of the echo envelope when tumor tissue regions becomes less dense and having a more consistent structure. Although tumor tissue is known to have a chaotic nature \citep{alk09}, the change in the tissue scatterer textural properties in a sense of becoming less random (e.g. inducing much less angiogenesis), or even the possible reduction in the number of effective scatterers in some parts of the tumor tissue (e.g. more necrotic regions) could be attributed to initial response to treatment and tumor becoming less aggressive. This explains why the Rayleigh distribution in Table 1 did not perform as well as the Nakagami distribution family which can take into account such variability in both tissue scatterer distribution and number density. On the other hand, it is not expected to encounter periodic patterns within the tumor tissue due to its chaotic nature, therefore dominance of periodic or coherent patterns in the modeled tissue texture is very rare, which explains the reduced classification performance based on the Rician distribution. As a result, the spatial scatterer distribution non-uniformities, such as clusterings in scatterers or regions with smaller effective numbers of scatterers cannot be effectively modeled. Also the K-distribution can be used to model the spacing or regularity in the spatial scatterer distribution, such as in partially developed speckle based on a low-average number of effective scatterers per resolution cell; however, it had the least classification performance in terms of predicting liver tumor response to chemotherapy treatment. This can be seen as tissue scatterers in liver tumors tend to have a high degree of number density which the model does not cover, and relying on characterizing the tissue randomness only might not suffice. Also the results in Table 1, 2, and 3 show the NIG recorded nearly similar performance in terms of classification accuracy to that of the Nakagami model. The main advantage of the extended NIG model over the bi-parameter Nakagami model is the better coverage of the coherent-diffuse region for low scatterer number density. Since there is not much obvious structural or periodic components in the tissue echo signal of the liver tumor, both models exhibited similar behavior. Moreover, the higher sensitivity values for the NIG model suggest that it is better in predicting response to chemotherapy treatment (i.e. aggressive tumors that are responding to treatment), while the Nakagami model is more precise and has the least overall FP rate.

Finally, the relatively low digital sampling rate could be a potential limitation. The effect of sparse sampling of the analog signal on the fractal analysis of the RF signal needs to be further investigated. Also as tumors may be localized at varying depths, this may decrease the amplitude of the RF data. Therefore the performance of the derived model parameters based on the fractal analysis technique as a function of depth could serve as future work. Furthermore, the study of the link between the FD derived from parametric images of weak scattering sources to the spatial distribution of tissue scatterers would assist in better understanding the robustness of the fractal analysis approach. 

\section*{Conclusions}
\label{Conclusions}
This paper presents an evaluation of different statistical models that cover different aspects of the tissue scatterers distributions and densities. The aim is to understand which model can give a better-fit of the complex nature of liver tumors by providing discriminative features for supporting clinical diagnosis as part of healthcare delivery. The tumor heterogeneity was assessed via its tissue fractal properties derived from the RF envelope-detected signal and used as features for assessing classification performance. The Nakagami distribution family was the most effective in characterizing the liver tumor tissue, with the simple bi-parametric model achieving equal accuracy and better precision as compared with the complex 4-parameter extended version. Moreover, fractal analysis of the echo signal was found to give additional information about the underlying tissue structure, and hence better indicates response to chemotherapy treatment. These findings suggest that the simple and general models of the Nakagami distribution family along with the fractal analysis of the RF parametric images have promising potential in providing improved ultrasound tissue characterization of liver tumors.

\section*{Acknowledgements}
\label{Ack}
This work was support by the Engineering and Physical Sciences Research Council and Wellcome Trust Grant WT 088877/z/09/z. The authors would like to thank the anonymous reviewers for their constructive comments and suggestions to improve the quality of the paper.





\pagebreak


\bibliographystyle{elsarticle-num}
\bibliography{UMB}

\end{document}